# Analysis of a capped carbon nanotube by linear-scaling density-functional theory

C. J. Edgcombe*[a], S. M. Masur[a], E. B. Linscott[b], J. A. J. Whaley-Baldwin[b] and C. H. W. Barnes[a]
[a]TFM Group, Department of Physics, University of Cambridge, CB3 0HE, Cambridge, UK
[b]TCM Group, Department of Physics, University of Cambridge, CB3 0HE, Cambridge, UK

**Abstract**

The apex region of a capped (5,5) carbon nanotube (CNT) has been modelled with the DFT package ONETEP, using boundary conditions provided by a classical calculation with a conducting surface in place of the CNT. Results from the DFT solution include the Fermi level and the physical distribution and energies of individual Kohn-Sham orbitals for the CNT tip. Application of an external electric field changes the orbital number of the highest occupied molecular orbital (the HOMO) and consequently changes the distribution of the HOMO on the CNT.

## 1. Introduction

The challenge of describing field-emitting structures and field emission continues to be addressed by development of methods of analysis. All these methods preserve the concept of tunnelling through a classically-forbidden region that provided one of the first successes of wave mechanics [1]. However, many further details of the emission process have proved difficult to describe simply, while estimation of practical parameters such as the emitting area remains a challenge for both experimental and theoretical techniques. Recent developments in microscopy, such as near-field scanning electron microscopy [2], have increased the need for methods that can describe the distribution of emitted current in detail. The widely-known semi-analytic theory uses classical modelling, with known limitations. However, density functional theory (DFT) can describe some behaviour from first principles with more accuracy, over a limited volume. In 2002, Han *et al* presented a study of a time-dependent (5,5) carbon nanotube (CNT) [3], and another study of a (10,10) CNT [4]. Zheng et al [5] divided the length of a SWCNT into subranges, iterating by solving each subrange in turn with DFT while the others were simulated by molecular mechanics. Csányi et al [6] introduced the use of non-plane-wave basis functions (as outlined below) and provision of the exterior boundary conditions for the DFT solution from an equivalent classical calculation. Wang *et al* [7] applied multi-scale DFT to model edges of graphene and open-ended CNTs, and Li [8] extended the method to project the current distribution to a screen.

Here we outline the program ONETEP [9], which is a linear-scaling code based on time-independent DFT, and give a survey of its results for a capped single-walled carbon nanotube (SWCNT). The initial motivation was that the use of plane waves as basis functions, as found in many DFT packages, is undesirable for non-periodic systems of hundreds of atoms or more. The number of plane waves and the solution time needed to model the system accurately would both be very large. As an

---

*corresponding author, *Email address:* cje1@cam.ac.uk

alternative, ONETEP (Order-N Electronic Total Energy Package) [9] uses a localised basis with which the solution time scales only as the number of electrons in the model. The method is usable at present only with external fields below the threshold for current flow, yet it has yielded a rich seam of data. Results outlined here, obtained both with zero external field and with applied field below threshold, include the spatial distribution of charge density, orbital energies at non-zero temperatures, distribution of the HOMO, distribution of total potential energy $E$ and energy due to non-classical effects ($E_{XC}$), and total and local densities of states. The calculated $E$ is compared with the classical potential energy (from work function and image) for the SWCNT.

## 2 Density functional theory

DFT takes advantage of the fact that the total energy of a many-body electron system is an (unknown) functional of the total electronic density $n(\mathbf{r})$ [10]. The Kohn-Sham construction replaces the intractable many-body electron problem with an auxiliary system of non-interacting electrons with an identical electronic density (and therefore identical total energy). The Kohn-Sham energy functional is given by

$$E[n] = \sum_i \langle\psi_i \left| -\frac{1}{2}\nabla^2 \right| \psi_i\rangle + E_{cores}[n] + E_{ext}[n] + E_{Coul}[n] + E_{XC}[n]$$

where the first term on the right represents the kinetic energy of a set of non-interacting Kohn-Sham orbitals $\psi_i$, the second is the energy due to the interaction of the electrons with the atomic cores, the third is the energy due to any externally applied field, the fourth is the Coulomb repulsion of the electron density and the fifth term – the "exchange-correlation functional" – accounts for the effects of exchange and correlation. In practice, $E_{XC}$ is not known exactly and is approximated. $E[n]$ is minimised by solving a Schrödinger-like equation for the Kohn-Sham orbitals subject to a density-dependent effective single-particle potential [11]. The work function is given by the negative of the HOMO energy (relative to vacuum energy), but because of the approximation used for $E_{XC}$ the energies of Kohn-Sham HOMOs are known to agree only moderately well with other estimates of work function for CNTs [12]. Detailed studies ([13], [14]) have discussed suitable forms to obtain better estimates of the orbital energies.

In ONETEP, the density is represented in a basis of non-orthogonal Wannier functions (NGWFs) $\{\phi_\alpha(\boldsymbol{r})\}$ and a kernel matrix $K^{\alpha\beta}$:

$$n(r) = \sum_\alpha \sum_\beta \phi_\alpha(\boldsymbol{r}) K^{\alpha\beta} \phi_\beta(\boldsymbol{r})$$

The solution method varies both the kernel coefficients and the Wannier functions adaptively to minimise the total energy. Because the NGWFs are spatially localised, resulting matrices such as the kernel are sparse and consequently the algorithm scales linearly with the number of electrons N. For problems with N of the order of hundreds, this property offers a great benefit relative to other methods whose solution time is proportional to $N^3$ or more. In the present calculation, the number of Kohn-Sham orbitals equals the number of electrons, so in the absence of applied field half these orbitals are occupied, with two electrons of opposite spins per orbital.

ONETEP typically uses a simulation cell with periodic boundary conditions, but it has been extended to handle aperiodic systems [6]. This is achieved by calculating $V_{Coul}[n] + V_{ext}[n]$ via the Poisson equation subject to fixed, user-specified boundary conditions. Here the potentials $V_i$ are related to the corresponding energy $E_i$ by $E_i = \int n(\boldsymbol{r})\, V_i(\boldsymbol{r})\, d\boldsymbol{r}$. In this case the charge density $n(\boldsymbol{r})$ is that due to the electronic density plus a smeared Gaussian representation of the ionic cores. The cores are included in this way for applications where the permittivity is not $\varepsilon_0$ (e.g. for implicit solvent models) in order to calculate accurately the electrostatic response of the dielectric. The contribution of the smeared ions is subsequently disentangled for calculating potentials and energies. For more details see [15] and [16]. In addition to imposing external boundary conditions on the electrostatic potential, the total system can be given a net charge (as found in the macroscopic system studied here).

The quantities that can be reported by ONETEP include charge density, potential energy, local density of states (LDoS) and spatial distribution of individual orbitals. These are reported at points on a regular cubic grid. In addition, the energy and occupation of orbitals and Fermi level are available. Calculations at non-zero temperatures are made by the ensemble-DFT method [17]. The program calculates the chemical potential and assumes the Fermi-Dirac distribution of orbitals. In this paper we quote energies relative to the Fermi level (treated here as identical to the chemical potential), with the warning that energy differences from this level may be underestimates [11].

**3. Method of calculation**

To obtain a solution within acceptable computing time, we define a limited volume (the DFT box) over which the density functional calculation will be solved. The volume used in the present work was a cube of side 5 nanometres. Values were calculated at points on a cubic grid at intervals of 12 picometres. Using the program Virtual Nanolab [18], we built a single-walled cylindrical (5,5) tube of carbon atoms, one end of which was capped with half a $C_{60}$ molecule (having a pentagon on its axis). The combination of tube and hemisphere contained 150 carbon atoms and was terminated at the uncapped end with 10 hydrogen atoms. The atomic positions were optimised until the force on each atom was less than 0.05 eV/Å. In the optimised CNT, the carbon cores are at a radius of 0.317 nm from the axis and form a structure 1.83 nm long (Fig. 1(b)).

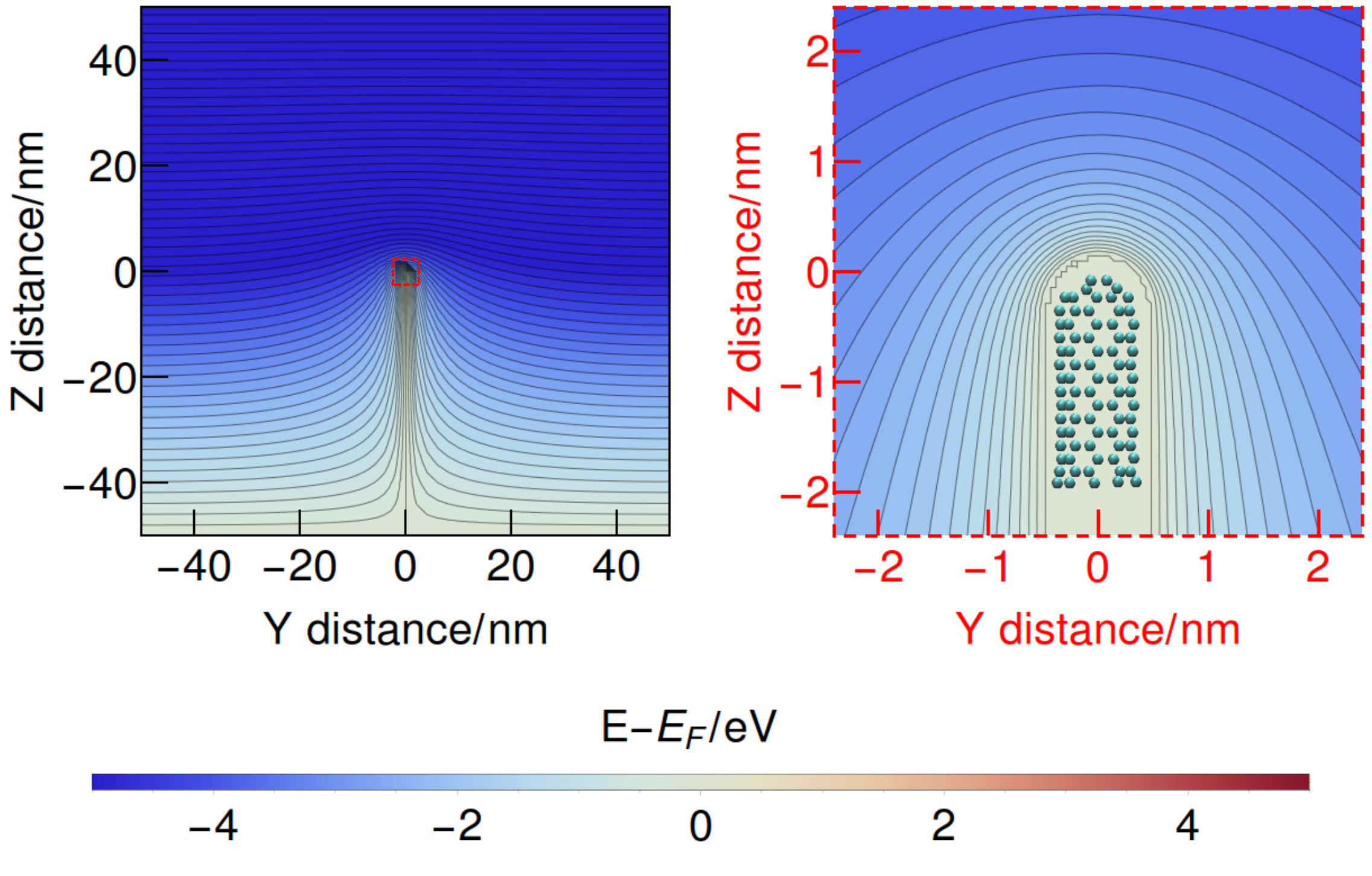

Fig. 1. Geometry analysed: (a) CNT supported on planar cathode electrode at z = –50 nm, with parallel anode plane at z = +50 nm. The small red box around the CNT tip defines the volume shown in (b); (b) the final 1.8 nm of the capped (cylindrical) CNT, in the (cubic) DFT box of side 5 nm. The equipotentials at intervals of 0.2 V are from the classical solution for a conductor with the diameter of the Fermi equipotential for the CNT, with an anode-cathode voltage of 5.4 V.

To solve the system with an applied electric field, we used a combination of classical and DFT calculations. The system for classical calculation consisted of parallel cathode and anode planes of diameter 100 nm and separated by 100 nm, with the CNT of total length 50.3 nm standing perpendicular to and in contact with the cathode plane (Fig. 1(a)). As the (5,5) CNT is known to be a conductor, it was modelled by a conducting rod with a hemispherical end and with diameter 0.978 nm, chosen to approximate that of the Fermi equipotential (as discussed in Section 5.2). The voltage distribution over the surface of the DFT box was found using the solver FlexPDE [19] and interpolated onto the mesh for the DFT box. This potential found by the classical calculation ignores the work function of the emitter (see Section 5.3). Although in the classical model the field is zero in the body of the rod, the DFT calculation solves for the varying potential energy distribution through the CNT.

Applying an external field induces additional charge onto the CNT. In the classical model, with the CNT connected to the cathode plane, the charge induced within the defined DFT box is a fixed multiple of the anode-cathode voltage. For the DFT calculation, in contrast, the short section of the CNT within the box is not connected to any electrode, so the charge on it can be varied independently and this also varies its mean potential relative to the boundary (by Gauss' theorem). The system to be analysed by DFT thus differs from the macroscopic system, because it requires the induced charge to be specified, in addition to the boundary potential. This charge can be found (again using Gauss' theorem) by integrating the classical normal electric field over the surface of the box. Thus to provide initial conditions for the DFT calculation, both the boundary potential and the excess charge in the box are needed, and so the distributions of both voltage and normal electric field over the surface of the DFT box are extracted from the classical solution.

The DFT calculation reported here used the LDA functional with the PW92 parameterisation [20]. All calculations had an energy cut-off of 1000 eV. There were four NGWFs on each carbon atom and one on each hydrogen atom. All NGWFs had cutoff radii of 12 Bohr. Projector augmented-waves from the GBRV dataset were used to represent the atomic cores [21][22]. The calculation produces potentials relative to the vacuum potential, which is well defined when the external field is zero. For non-zero applied fields the potential reference is less clear. We can however use the rule known for semiconducting systems, that the macroscopically measurable voltage of an electrode corresponds to the Fermi level for that electrode. We assume the CNT to be in a simple circuit with cathode and anode as described above, in which the voltage of the cathode and CNT is held fixed as the anode-cathode voltage is varied. Then the Fermi level of the CNT remains constant as the external field varies. Thus, provided we can identify the Fermi levels for different applied fields, we can compare the potential distributions around the tip by aligning their Fermi levels. When this is done, the potential distribution near the atomic cores is found to vary very little with applied field.

## 4. Results for capped (5,5) SWCNT

### *4.1 Classical solution with applied field*

The lowest non-zero anode voltage considered here is 5.4 V relative to the cathode, corresponding to a background field (in the absence of the nanotube) of 0.054 $Vnm^{-1}$. The resulting equipotentials for this system are shown in Fig. 1(a). Fig. 1(b) shows a section through the volume of the DFT box, with classical equipotentials for voltages greater than the cathode voltage of 0V and with the equilibrium structure of the carbon cores superimposed.

### *4.2 DFT solution for charge density*

The electron density distribution found by DFT with zero applied field is shown in Fig. 2. Isosurfaces are plotted for densities of 0.02%, 0.25% and 1% of the global maximum density (717 $eÅ^{-3}$ at zero field). Most of the charge is localised near the carbon cores, but there is also a continuous sheath of relatively low density at a slightly greater distance from the cores. This continuous sheath provides a conducting path for charge to move along and around the CNT surface, up to a density between 0.25% and 1% of the global maximum.

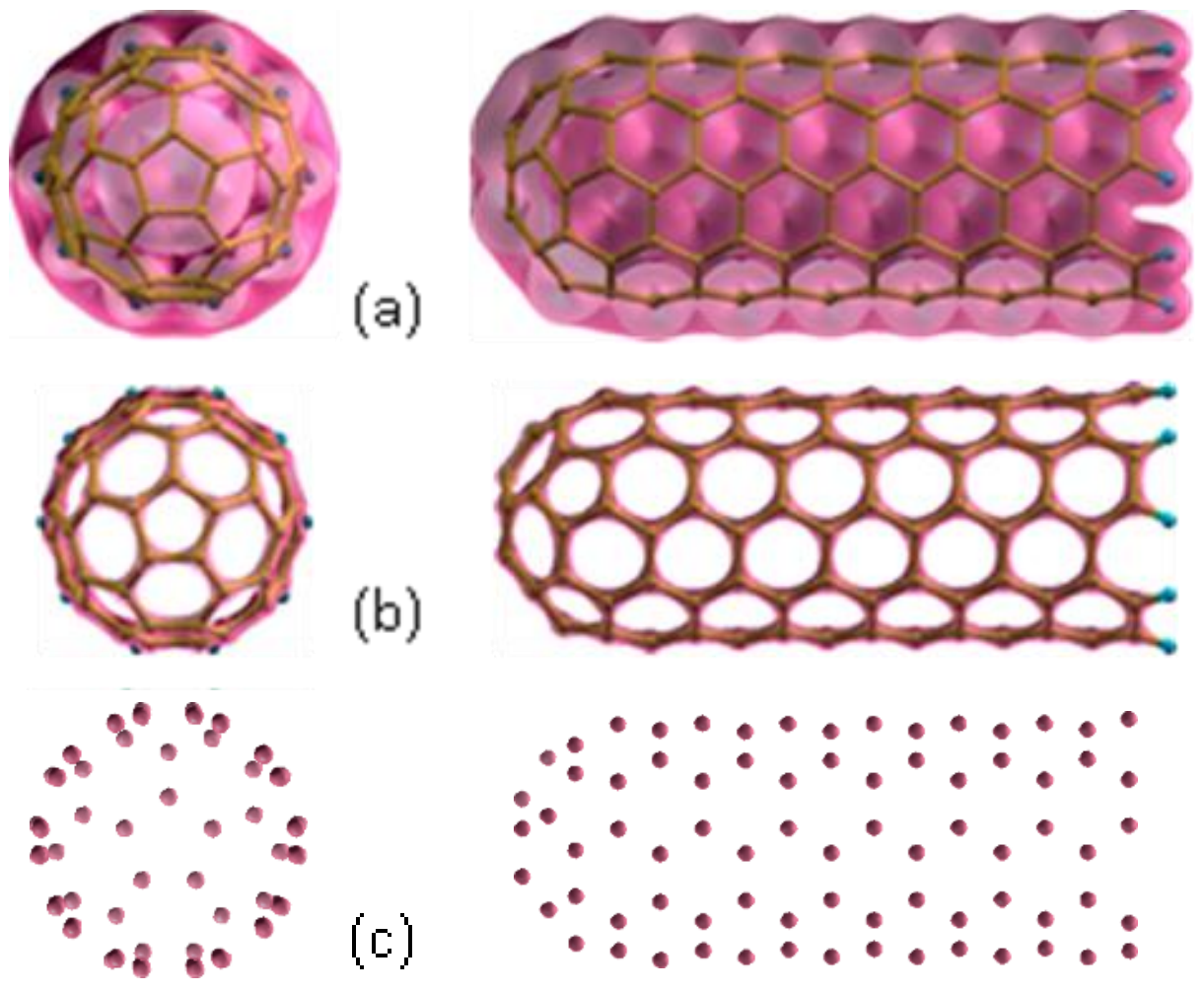

Fig. 2. Electron density isosurfaces for zero applied field (those with 0.054 V/nm applied field are identical in appearance at this scale). The isosurfaces are at (a) 0.02%, (b) 0.25%, (c) 1.00% of the global maximum charge density (≈ 0.14, 1.8, 7.2 $eÅ^{-3}$ respectively). In (a) and (b), isosurfaces are in purple with the atomic frame of the CNT added. The views on the left are from the cap end.

A background field of 0.162 $Vnm^{-1}$ induces three additional electrons into the DFT box. The corresponding change of charge density is shown in Fig. 3. The magnitude of the change in density shown by these isosurfaces is $5 \times 10^{-3}$ $eÅ^{-3}$, which is about $3 \times 10^{-3}$ of the static charge density of the middle surface in Fig. 2.

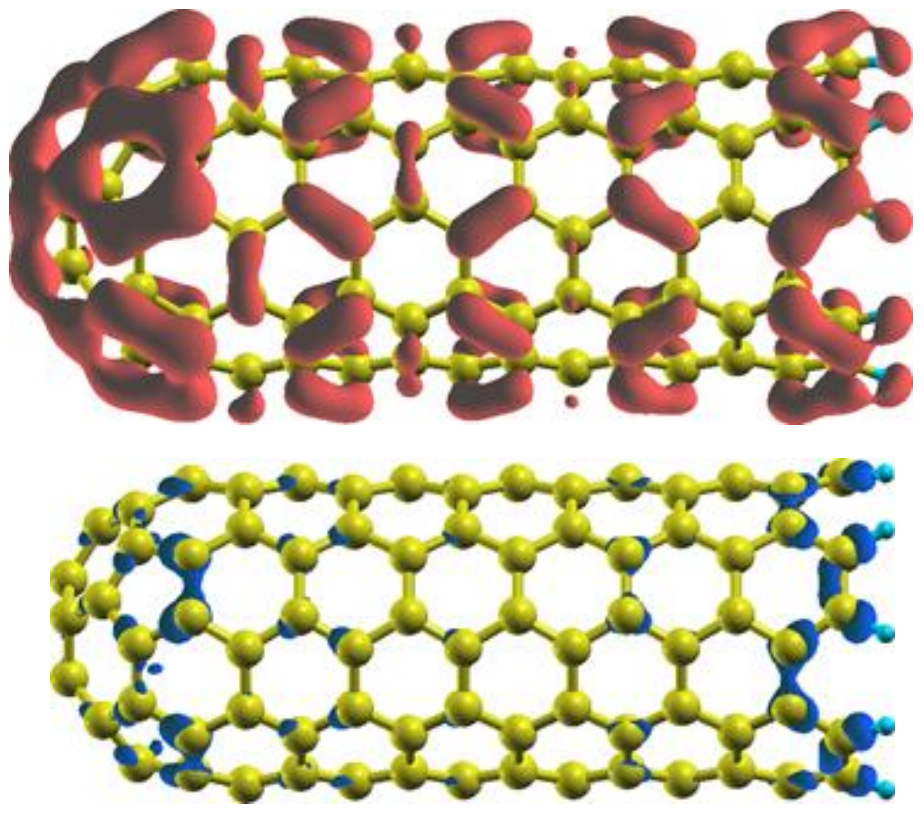

Fig 3. Change in electron density on application of 0.162 $Vnm^{-1}$ with three added electrons. The red isosurface shows where the electron density has increased by $5 \times 10^{-3}$ $eÅ^{-3}$ and the blue isosurface shows where the density has decreased by $5 \times 10^{-3}$ $eA^{-3}$, relative to zero-field values.

### *4.3 Local density of states (LDoS)*

The LDoS for the four planes of carbon atoms perpendicular to the CNT axis and closest to the end cap has been plotted for four values of applied field. Applying a field produces small changes in the pattern but the general forms remain similar. The general pattern, relative to the Fermi level, moves with applied field, but the apparent movement depends on whether the Fermi level used is always that for zero field or as determined for each applied field. The behaviour relative to the Fermi level for each applied field is shown in Fig. 4.

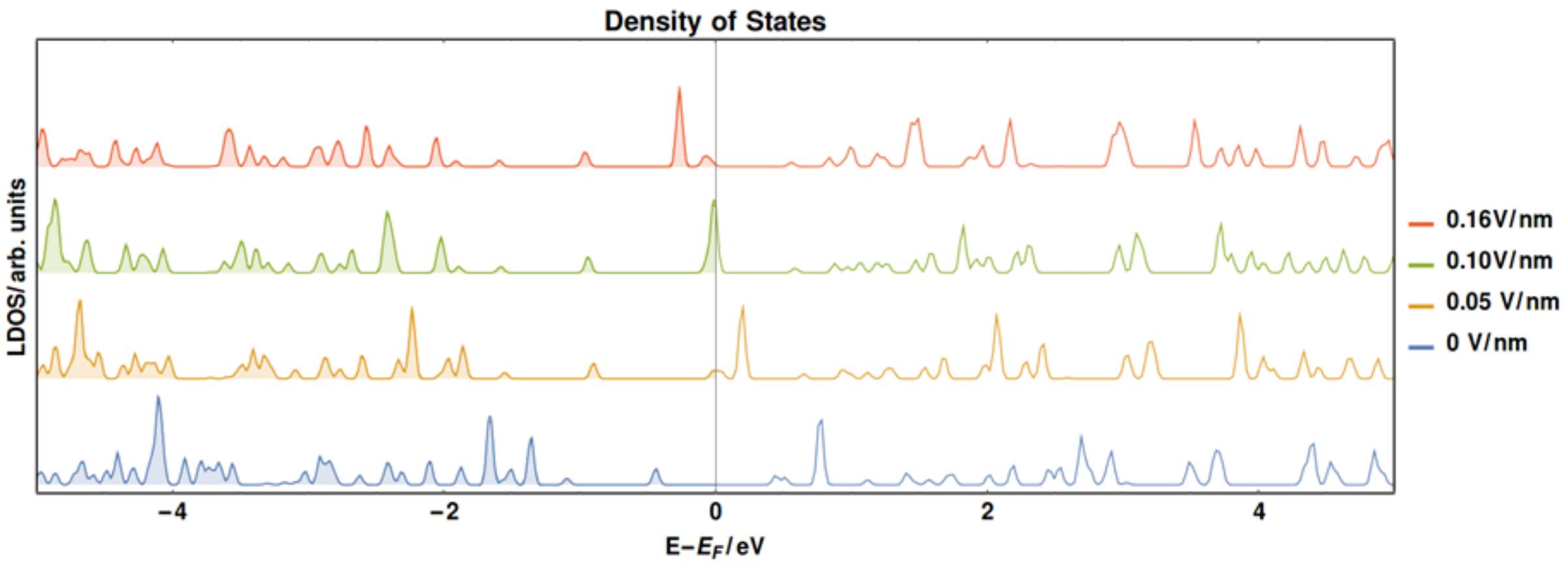

Fig. 4 Local density of states as a function of energy near the Fermi level, for four values of applied field. The density has been collected from the four planes of carbon atoms perpendicular to the axis and closest to the end cap. The delta-functions have been broadened by Gaussians corresponding to a temperature of 300K.

### *4.4 Spatial distribution of the HOMO*

Fig. 5 shows isosurfaces for one amplitude (+ and –) of the HOMOs found with four external fields. For larger amplitudes, the distribution appears discontinuous, like the charge density in Fig. 2(c). The four fields used for these plots add electrons into the DFT box successively with alternate spins. Each orbital can accept two spins, so the HOMO number increases by one after two electrons are added. The distribution on the CNT surface thus changes with the applied field.

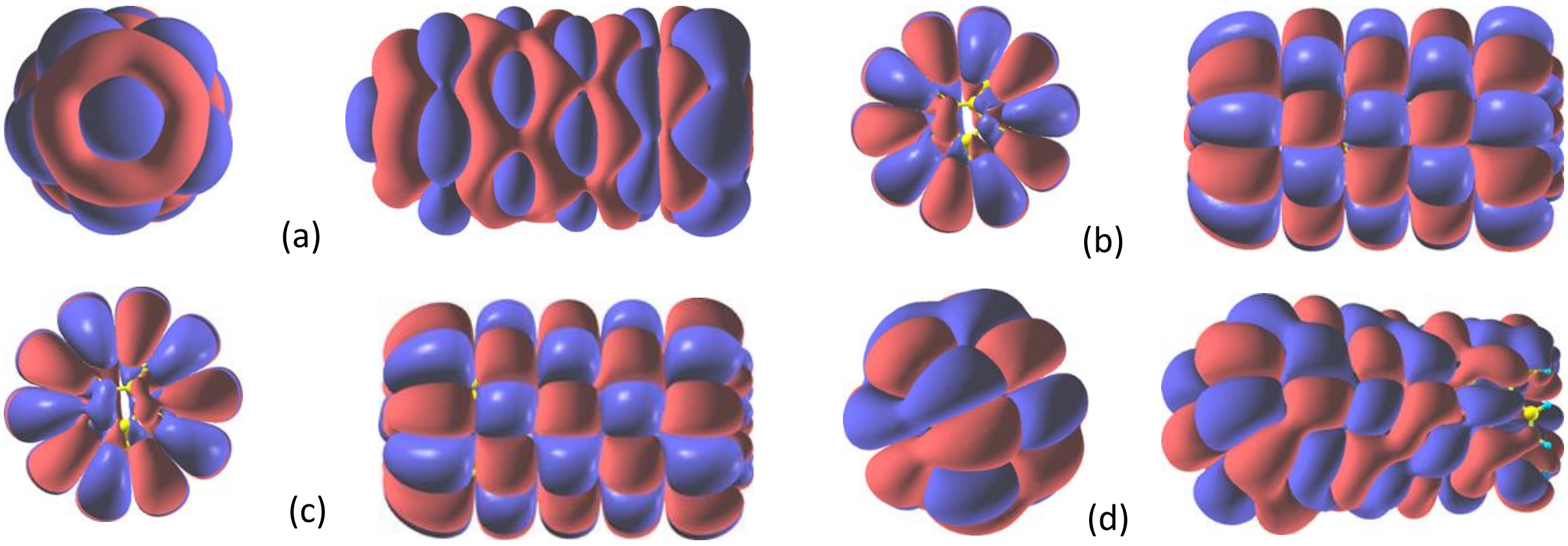

Fig. 5 HOMO isosurfaces at amplitude of ±0.0004 $(eA^{-3})^{1/2}$, with Kohn-Sham orbital numbers $n_{KS}$: (a) in zero field, $n_{KS}$ = 305↑ and 305↓; (b) in 0.054 $Vnm^{-1}$, 306↑ and 305↓; (c) in 0.108 $Vnm^{-1}$, $n_{KS}$ = 306↑ and 306↓; (d) in 0.162 $Vnm^{-1}$, $n_{KS}$ = 307↑ and 306↓. Only the ↑ spin of each pair is plotted here.

### *4.5 Potential with applied field*

A longitudinal section through the DFT box, with equipotentials resulting from an applied anode-cathode voltage of 8V, is shown in Fig. 6. The distribution of potential near the hemispherical cap and cylindrical surface is much smoother than that of charge density, which is reasonable since the potential is a double integral of the charge density, in accordance with Poisson's equation. The applied field was lower than the threshold value for emission and, at this low field, no indication has been seen of field penetration at the apex. The radius of the Fermi equipotential from the axis is 0.470 nm in zero field, and 0.49 nm in the background field of 0.054 Vnm$^{-1}$.

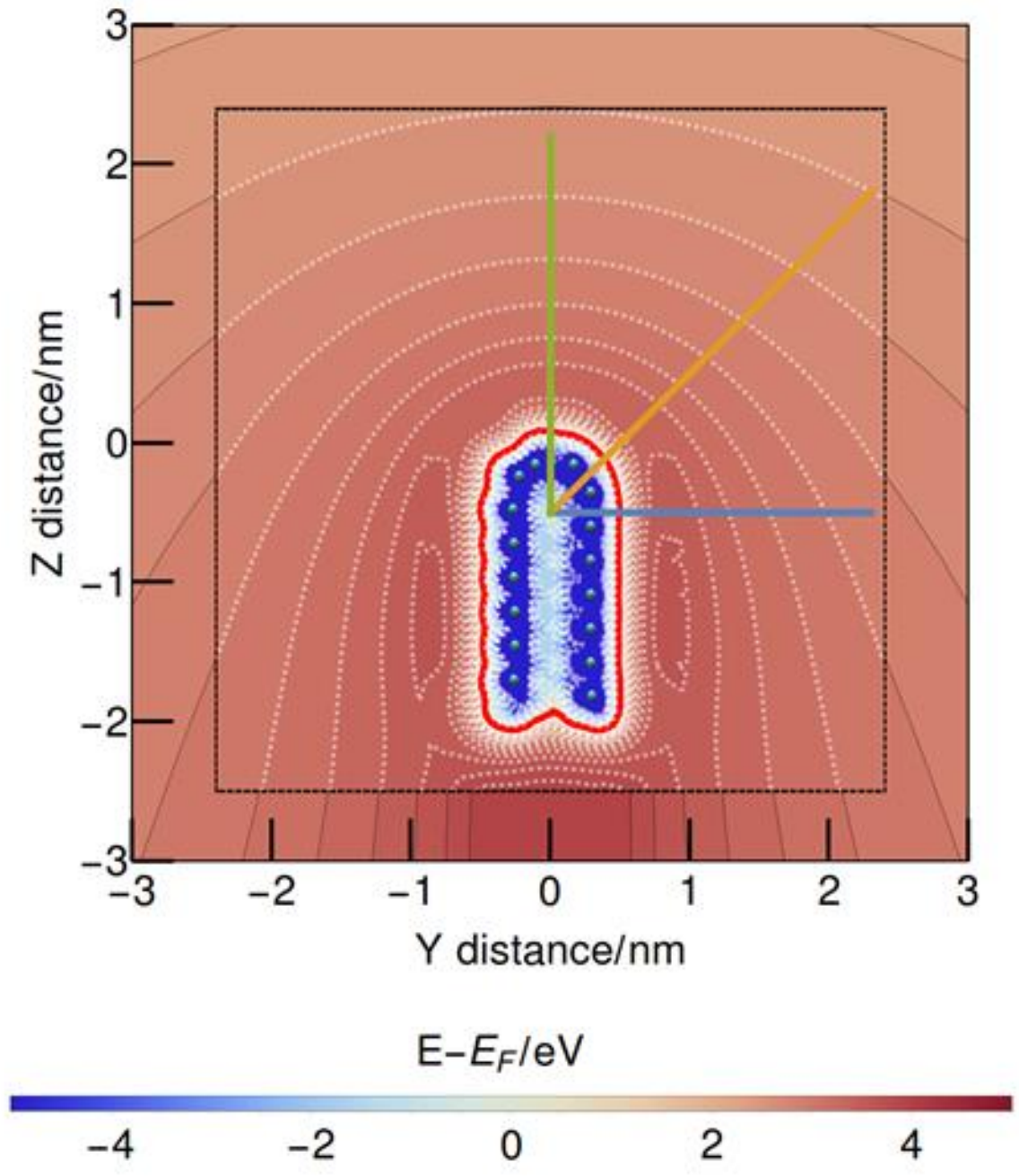

Fig. 6. Section containing the CNT axis, showing the DFT potential energy $E$ within the box for a background field of 0.054 Vnm$^{-1}$ and with 1 excess electron. The red outline is the Fermi equipotential ($E_F$). Equipotentials (in white) above and below $E_F$ are at intervals of 0.2 eV and 1 eV respectively. The coloured lines starting from the point ($z$ = –0.5nm, $r$= 0) indicate the directions (green = 0°, orange = 45° and blue = 90° relative to the axis) used for the plots of Fig. 7. The dots indicate the positions of carbon cores. The DFT potential is superimposed on part of the classical potential (raised by the work function), to show the agreement of the potential distributions on the interface.

The potential distribution (including $E_{XC}$) along the straight lines in Fig. 6 from the point ($z$ = –0.5nm, $r$ = 0), 0.1nm below the centre of the hemispherical cap, is shown in Fig. 7, relative to the Fermi level, for three directions relative to the axis and four values of background field: zero, 0.054, 0.108 and 0.162 Vnm$^{-1}$. For the conditions modelled in Fig. 7, it appears that a further increase in applied field above 0.162 Vnm$^{-1}$ would reduce the maximum potential on axis (plot (a)) to below the Fermi level, so causing strong emission.

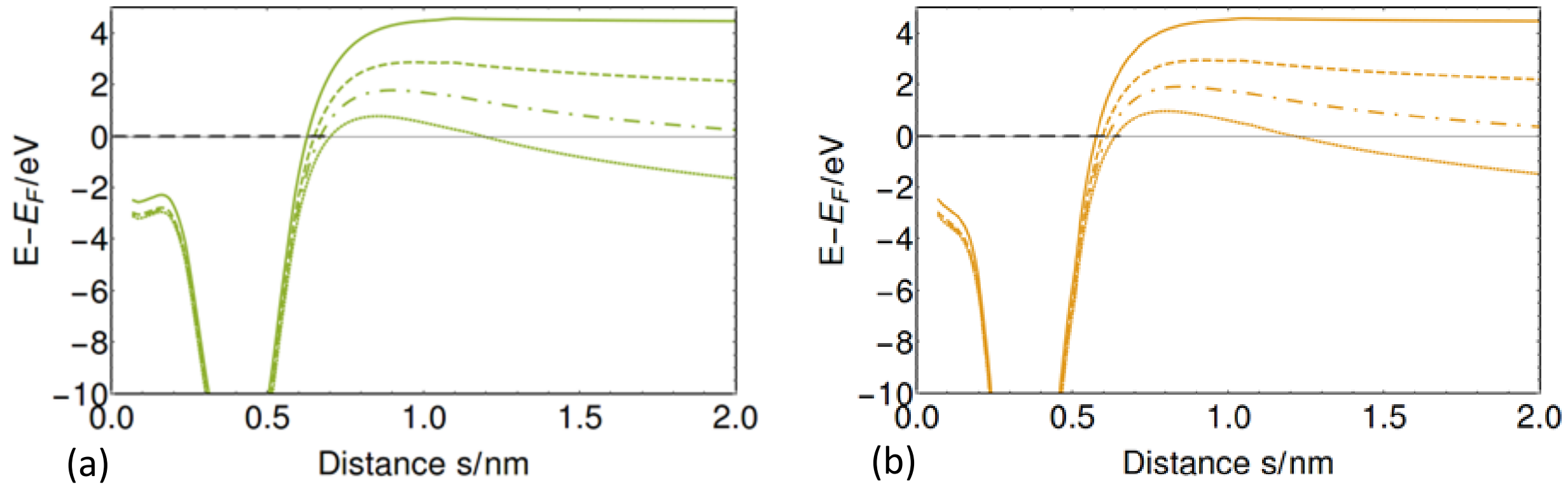

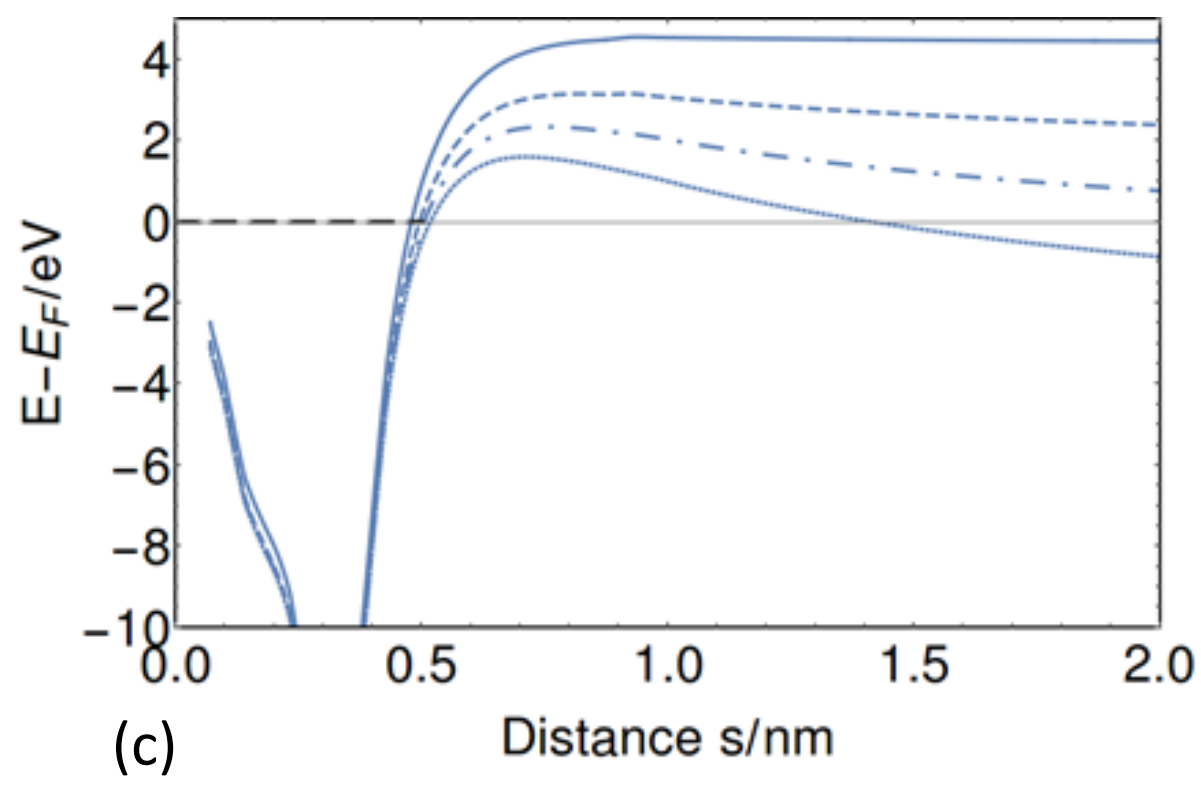

Fig. 7. Potential energy at 100 K, as function of distance $s$ from ($z$ = –0.5 nm, $r$ = 0), at angles relative to the axis of (a) zero; (b) 45 degrees; (c) 90 degrees, for four values of applied field: solid, zero field; dashed, 0.054 Vnm$^{-1}$; dot-dashed, 0.108 Vnm$^{-1}$; dotted, 0.162 Vnm$^{-1}$.

Fig. 8 shows more detail of the potentials ($E - E_{XC}$) and $E$ on the axis. The potential variation near the core includes the Coulomb term (due to valence electrons) and the core term (due to the nuclei and the inner electrons) included via the projector augmented-waves. The classical part of the potential ($E - E_{XC}$) shows a drop of a few eV in the core region, but the contribution from $E_{XC}$ produces a much greater drop in potential over the regions around the cores. The relation of the classical model to these potentials is discussed in Section 5.3.

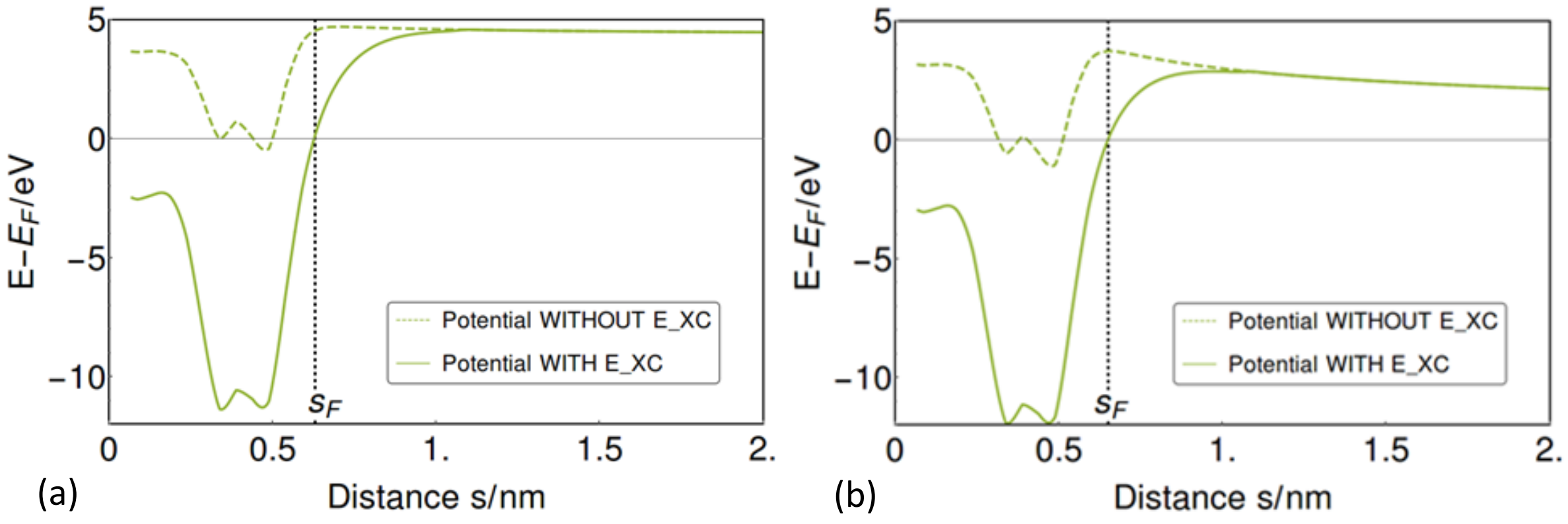

Fig. 8 Potential energy at 100 K as function of distance $s$ along axis from ($z$ = –0.5nm, $r$ = 0) for (a) zero applied field and (b) applied field of 0.054 Vnm$^{-1}$. Each plot shows two curves from a single calculation which included exchange and correlation effects; the lower curve is the total energy $E$ including $E_{XC}$, the other is the 'non-interacting' component ($E - E_{XC}$).

## 5. Discussion

The following remarks apply to the system modelled here, but not necessarily to other configurations of field emitters. It will be convenient to denote by $s_F$ the intersection of the axis with the Fermi equipotential of $E$, at zero external field, as shown at $s$ = 0.63 nm in Fig. 8(a).

### *5.1 Charge density*

Fig. 2 shows that much of the electron density is localised around the carbon cores, but there is also a continuous sheath of lower density that allows conduction over all the CNT surface. Any external electric field induces charge on this outer sheath. This screens the cores so that they are not much affected by the external field, as shown by the plots of potential in the core regions of Fig. 8.

### *5.2 The equivalent classical surface*

As suggested in Section 3 above, consistency between the classical and atomic models requires agreement as to what outline in the atomic model should be taken as 'the surface' of the equivalent classical conductor. As, in classical terms, the surface of a conductor is an equipotential, it should agree with an equipotential in the atomic model. Where details are available at atomic scale, an obvious possibility is to define the classical surface as the equipotential for the Fermi level at zero external field, as shown in Fig. 8 at $s_F$ . When field is applied, the Fermi equipotential moves away from $s_F$ and electrons are also able to move further from the cores.

### *5.3 Comparison of exchange and image potentials*

The upper traces in Fig. 8(a) and (b) show the energy due to the 'non-interacting' part of the DFT calculation (with $E_{XC}$ excluded). Outside the Fermi equipotential (that is, for $s > s_F$), these results from DFT are equivalent to the sum of the work function and the potential in applied field as calculated by semi-classical field-emitter theory. In that theory, before any image is considered, the work function is assumed to appear as a step function at the surface of the emitter.

Following the initial suggestion of Fowler and Nordheim [1], most semi-classical theory has used an image potential to describe the transition between the Fermi and vacuum levels. Fig. 8(a) shows that just outside the surface ($s > s_F$), the transit between these levels is not visible in the classical part of the DFT energy ($E - E_{XC}$) but does appear in the total energy $E$. The detailed DFT calculation shows that this transit is part of a larger change in $E$ that is continuous between the cores and vacuum and is due mainly to the strong core-directed field associated with $E_{XC}$. The upper part of this range of $E$, from the Fermi level to vacuum, is observed as the work function.

$E$ as calculated by ONETEP is plotted in Fig. 9 as a function of axial distance, for zero applied field and at 100 K. The equivalent classical models consisting of the work function plus an image potential for a planar or a spherical conductor are also shown. For each image calculation, there is the question of where to locate the conducting surface that provides an electrostatic mirror. Fig. 2(a) and (c) show that charge density is relatively weak between the cores, but a continuous shell of charge exists out to near the Fermi equipotential. We therefore choose the Fermi equipotential as the notional mirror surface for image calculations and approximate it by a plane or by a spherical surface whose radius is that of the Fermi equipotential at the axis. These approximating surfaces meet the axis at $s_F$, as defined in Sec. 5.2 and shown in Fig. 9.

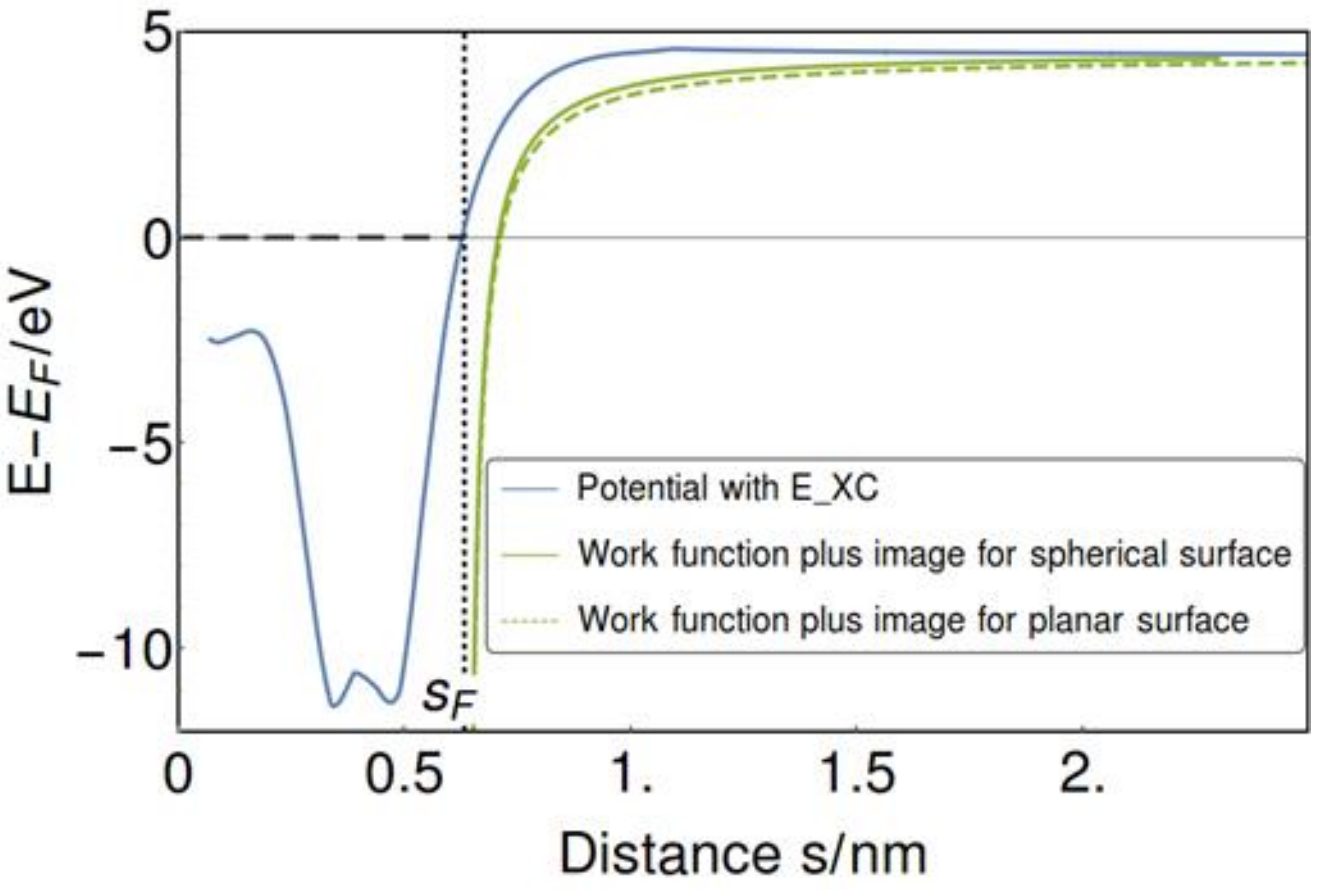

Fig. 9. Potential including $E_{XC}$ (in blue), as a function of distance along the axis. The cores are nearest to the axis at $s$ = 0.42 nm. Classical (work function + image) potentials are also shown (in green), for a plane conductor and for a spherical one with the radius of the Fermi equipotential, both meeting the axis at $s_F$.

It is known that $E_{XC}$ far outside a solid surface is given by the classical image potential [23],[24]. However, the asymptotic behaviour of $E_{XC}$ computed in practice depends on the choice of exchange-correlation functional. In Fig. 9, as $s$ increases beyond $s_F$, $E$ appears to fall off exponentially, but this is a known property of the LDA functional used [23],[25]. To improve agreement, it seems desirable to re-calculate using a functional with the desired asymptotic behaviour, such as the weighted density approximation (WDA) [23],[25].

## 6. Conclusions

Modelling the properties of a capped (5,5) CNT by linear-scaling DFT has given detailed estimates of the atomic properties that are not available from classical approximations. The available results include (a) the spatial distribution of charge density, showing a low-density sheath which provides conduction and screens the core potential from external fields; (b) the distribution and relative energies of individual orbitals including the HOMO; (c) the local density of states; (d) the distribution of total potential and also the estimated component due to exchange and correlation effects ($E_{XC}$). Results can be compared for zero external field and for applied fields that are below the threshold for electron emission. These results are obtained by use of the linear-scaling DFT program ONETEP over the apex region of the CNT, with boundary conditions set by classical modelling of the macroscopic system.

Results by this method show that on the axis near the cores, $|E_{XC}|$ is as large as 10 eV or more for the structure considered. The charge density around the cores is contained within the equipotential at the Fermi level; this equipotential in zero external field appears to be a suitable definition of 'the surface' for classical modelling.  Immediately outside the Fermi equipotential, the tail of $E_{xc}$ is the chief contribution to the change of $E$ between Fermi and vacuum levels that is known as the work function.

The spatial distribution of the HOMO at zero or low field is somewhat similar to that of the electronic charge.  Applying a field (below the emission threshold) induces unbalanced charge onto the CNT which changes the orbital number of the Kohn-Sham HOMO and consequently changes the distribution of the HOMO on the CNT.

**Acknowledgments**
The authors thank Prof. M.C. Payne for the opportunity to use ONETEP, and Prof. C.-K. Skylaris, Dr J. Dziedzic, and Dr. G. Constantinescu of the TCM Group of C.U. Department of Physics for their advice on running it.

Figures 1 and 6 were plotted using Mathematica [26].  Figures 2, 3 and 5 were plotted using XCrySDen [27].

The authors declare no conflicting interests.  The research leading to these results has received funding from the People Programme (Marie Curie Actions) of the European Union's Seventh Framework Programme FP7/2007-2013/ under REA grant agreement n 606988.

**References**
[1] R.H. Fowler, L. Nordheim, Electron Emission in Intense Electric Fields, Proc. R. Soc. London A-Mathematical Phys. Sci. 119 (1928) 173–181. doi:10.1098/rspa.1928.0091.
[2] D.A. Zanin, H. Cabrera, L.G. De Pietro, M. Pikulski, M. Goldmann, U. Ramsperger, D. Pescia, J.P. Xanthakis, Fundamental Aspects of Near-Field Emission Scanning Electron Microscopy, in: Adv. Imaging Electron Phys. Vol 170, 2012: p. 227.
[3] S. Han, M.H. Lee, J. Ihm, Dynamical simulation of field emission in nanostructures, Phys. Rev. B - Condens. Matter Mater. Phys. 65 (2002) 085405. doi:10.1103/PhysRevB.65.085405.
[4] S. Han, J. Ihm, First-principles study of field emission of carbon nanotubes, Phys. Rev. B - Condens. Matter Mater. Phys. 66 (2002) 241402. doi:10.1103/PhysRevB.66.241402.
[5] X. Zheng, G.H. Chen, Z. Li, S. Deng, N. Xu, Quantum-Mechanical Investigation of Field-Emission Mechanism of a Micrometer-Long Single-Walled Carbon Nanotube, Phys. Rev. Lett. 92 (2004) 12–15. doi:10.1103/PhysRevLett.92.106803.
[6] G. Csányi, C.-K. Skylaris, A. Nevidomskyy, C.J. Edgcombe, Ab initio approach to a lightning rod, 30 November 2005. http://www.tcm.phy.cam.ac.uk/~ndd21/esdg.html (accessed 28 Dec 2018).
[7] W. Wang, J. Shao, Z. Li, The exchange–correlation potential correction to the vacuum potential barrier of graphene edge, Chem. Phys. Lett. 522 (2012) 83–85. doi:doi:10.1016/j.cplett.2011.12.002.
[8] Z. Li, Density functional theory for field emission from carbon nano-structures, Ultramicroscopy. 159 (2015) 162–172. doi:10.1016/j.ultramic.2015.02.012.
[9] C.-K. Skylaris, P.D. Haynes, A.A. Mostofi, M.C. Payne, Introducing ONETEP: Linear-scaling density functional simulations on parallel computers, J Chem Phys. 122 (2005) 084119.

doi:10.1063/1.1839852.
[10] P. Hohenberg, W. Kohn, The Inhomogeneous Electron Gas, Phys. Rev. 136 (1964) B864. doi:10.1103/PhysRev.136B864.
[11] W. Kohn, L.J. Sham, Self-Consistent Equations Including Exchange and Correlation Effects, Phys. Rev. 140 (1965) A1133–A1138.
[12] R. Stowasser, R. Hoffmann, What Do the Kohn - Sham Orbitals and Eigenvalues Mean ?, J. Am. Chem. Soc. (1999) 3414–3420.
[13] S. Hamel, P. Duffy, M.E. Casida, D.R. Salahub, Kohn – Sham orbitals and orbital energies : fictitious constructs but good approximations all the same, J. Electron Spectrosc. 123 (2002) 345–363.
[14] T. Grabo, E.K.U. Gross, Density-functional theory using an optimized exchange-correlation potential, Chem. Phys. Lett. 240 (1995) 141–150.
[15] J. Dziedzic, H.H. Helal, C.-K. Skylaris, A.A. Mostofi, M.C. Payne, Minimal parameter implicit solvent model for ab initio electronic-structure calculations, Europhys. Lett. 95 (2011) 43001. doi:10.1209/0295-5075/95/43001.
[16] J. Dziedzic, S.J. Fox, T. Fox, C.S. Tautermann, C. Skylaris, Large-scale DFT calculations in implicit solvent—A case study on the T4 lysozyme L99A/M102Q protein, Int. J. Quantum Chem. 113 (2013) 771–785. doi:10.1002/qua.24075.
[17] N. Marzari, D. Vanderbilt, M.C. Payne, Ensemble density-functional theory for ab initio molecular dynamics of metals and finite-temperature insulators, Phys. Rev. Lett. 79 (1997) 1337–1340. doi:10.1103/PhysRevLett.79.1337.
[18] J. Schneider, J. Hamaekers, S. Chill, S. Smidstrup, J. Bulin, R. Thesen, A. Blom, K. Stokbro, ATK-ForceField: a new generation molecular dynamics software package, Model. Simul. Mater. Sci. Eng. 25 (2017) 085007. doi:10.1088/1361-651X/aa8ff0.
[19] PDE Solutions Inc, FlexPDE, (n.d.). https://www.pdesolutions.com.
[20] J.P. Perdew, Y. Wang, &ccnrate and simple analytic representation of the electron-gas correlation energy, Phys. Rev. B. 45 (1992) 13244–9. doi:10.1103/PhysRevB.45.13244.
[21] P.E. Blöchl, Projector augmented-wave method, Phys. Rev. B. 50 (1994) 17953–17979. doi:10.1103/PhysRevB.50.17953.
[22] K.F. Garrity, J.W. Bennett, K.M. Rabe, D. Vanderbilt, GBRV high-throughput pseudopotentials, (2015). doi:https://doi.org/10.1016/j.commatsci.2013.08.053.
[23] O. Gunnarsson, M. Jonson, B.I. Lundqvist, Descriptions of exchange and correlation effects in inhomogeneous electron systems, Phys. Rev. B. 20 (1979) 3136–3164. doi:10.1103/PhysRevB.20.3136.
[24] C.O. Almbladh, U. Von Barth, Exact results for the charge and spin densities, exchange-correlation potentials, and density-functional eigenvalues, Phys. Rev. B. 31 (1985) 3231–3244. doi:10.1103/PhysRevB.31.3231.
[25] P. García-González, J.E. Alvarellos, E. Chacón, P. Tarazona, Image potential and the exchange-correlation weighted density approximation functional, Phys. Rev. B - Condens. Matter Mater. Phys. 62 (2000) 16063–16068. doi:10.1103/PhysRevB.62.16063.
[26] Wolfram Research, Mathematica, (2018).
[27] A. Kokalj, Computer graphics and graphical user interfaces as tools in simulations of matter at the atomic scale, Comput. Mater. Sci. 28 (2003) 155–168. doi:10.1016/S0927-0256(03)00104-6.